\documentstyle[11pt,paspconf,epsf]{article}

\begin{document}

 \title{Influence of Hipparcos on Hyades age estimates 
from three binary systems}

\author{E. Lastennet}
 \affil{Astronomy Unit,  Queen  Mary and  Westfield  College, Mile End
 Road, London E1 4NS, UK}
    
\author{D. Valls-Gabaud}
 \affil{UMR  CNRS   7550, Observatoire    Astronomique,  11,  rue   de
 l'Universit\'e, 67000 Strasbourg, France}
    
\author{Th. Lejeune}
 \affil{Astronomisches Institut der  Universit\"at Basel, Venusstr. 7,
 CH-4102 Binningen, Switzerland}
    
\author{E. Oblak}
 \affil{Observatoire de Besan\c  con, 41 bis avenue de l'Observatoire,
 F-25010 Besan\c con, Cedex, France}

\begin{abstract}

Three independent sets of stellar theoretical models are tested with 
well-detached systems of the Hyades open cluster: 51 Tau, V818 Tau, 
and $\theta^2$ Tau. The choice of these objects is discussed and a 
statistical method is described and applied to the colour-magnitude
diagram (CMD) of the selected stars, giving rise to contour 
levels in the metallicity-age plane. The effects of the Hipparcos 
parallaxes on these confidence regions, with influence on both age and 
metallicity, are studied for the V818 Tau system through a comparison 
with very accurate but older orbital parallaxes.
Finally, theoretical simultaneous age-metallicity estimates are given 
for the three binaries and compared with observational constraints. 

\end{abstract}


\keywords{binary systems, open cluster: Hyades}

\section{Introduction}

The main advantage to choose binaries which are members of the Hyades 
open cluster is that the heavy element abundance of the Hyades has been 
extensively studied and estimated by different authors.  
The more recent determinations of the Hyades metallicity have been reviewed 
by Perryman et al. (1998): [Fe/H] $=$ 0.14 $\pm$ 0.05, i.e. 
Z $=$ $0.024^{+0.0025}_{-0.003}$ assuming a subsolar helium abundance 
(Y $=$ 0.26), which is the value found by Lebreton et al. (1997) 
in order to reproduce the Hyades main sequence. 
A second advantage is that very accurate parallaxes are now available 
from Hipparcos for the systems 51 Tau, V818 Tau, and $\theta^2$ Tau.    
Moreover, Torres et al., 1997 ([TSL97a], [TSL97b]  and  [TSL97c]) 
obtained the first complete  visual-spectroscopic solutions for the 3
above-mention\-ned systems,  from  which  they carefully derived  very
accurate parallaxes and individual  masses.  They also gathered   some
individual photometric data  in  the Johnson system.  Furthermore,  we
found useful trigonometric   parallaxes information  in the  Hipparcos
catalogue (ESA, 1997).  By combining   the two  sources of data,    we
investigate  the influence of the  Hipparcos parallaxes  on our method
which  was  developed  to test   stellar  evolutionary  models  in  HR
diagrams.  

\subsubsection{Tests in the CMD : }
The tests we want to perform are the following :
\begin{enumerate}
\item 
to  check whether the two  components of the  systems  are on the same
isochrone, i.e. on   a  line defined by   the  same age and  the  same
chemical composition for the two single stars.
\item 
since all the  selected    stars are  members  of the  Hyades  whose
metallicity  has   been  well measured, we  can also check that the 
predicted metallicities from theoretical models are correct.
\item 
for 51 Tau and $\theta^2$ Tau, the individual stellar masses are known
with an accuracy of about 10\%, and for V818 Tau, masses and radii are
known with an accuracy close to 1-2\%, allowing further tests with the
theoretical models (see Lastennet et al. 1999b for details on this point).
\end{enumerate}

Therefore, if one  of these criteria  is  not clearly fullfilled  by a
given set  of  tracks, then these  models have  obvious problems because 
they do not account for several observational constraints (the
metallicity, mass, radius, and/or the photometric data).

\section{An example: the V818 Tau binary system}

The V818 Tau system is a double-lined eclipsing binary (Mc Clure, 1982) 
with very well estimated masses (actually the most accurate masses known 
for Hyades members). 
Indeed, the relative errors on the masses are less than 1\%, and the 
secondary component is particularly interesting because it is one of 
the rare stars which is less massive than the Sun and whose mass is 
known with such an accuracy (cf. Andersen 1991).

\subsection{Brief description of the method } 
In order to derive simultaneously the metallicity  (Z) and the age (t)
of the system,  and to produce confidence  level  contours (see Figure
1), we minimize the $\chi^2$-functional defined as:

\small
\begin{eqnarray}
 \chi^2 (t, Z) & = & \sum_{i=A}^{B} \left[ \left(\frac{\rm M_V(i)_{\rm
 mod}   - M_V(i)}   {\sigma(\rm  M_V(i))}\right)^2 +   \left(\frac{\rm
 (B-V)(i)_{\rm mod} - (B-V)(i)}{\sigma(\rm (B-V)(i))}\right)^2 \right]
\end{eqnarray}
\normalsize

where $A$ is the  primary and $B$ the secondary component.  M$_V$ and
(B$-$V) are the observed  values, and  M$_V$$_{\rm mod}$ and
(B$-$V)$_{\rm mod}$ are  obtained  from the synthetic  computations of
the BaSeL (Basel Stellar Library) models (see Lejeune et al. 1997, 
1998 and  Lastennet et al. 1999a) using a given set of stellar tracks 
from the Geneva group (see Charbonnel et al. 1993 and references therein), 
the Padova group (see Fagotto et al. 1994 and references therein), and 
from Claret \& Gim\'enez (1992) (CG92 thereafter). For reasons developed 
in Lastennet et al. (1999b), we assume that the calibrations from the 
BaSeL models are reliable enough for this work.

\begin{figure}
\plotfiddle{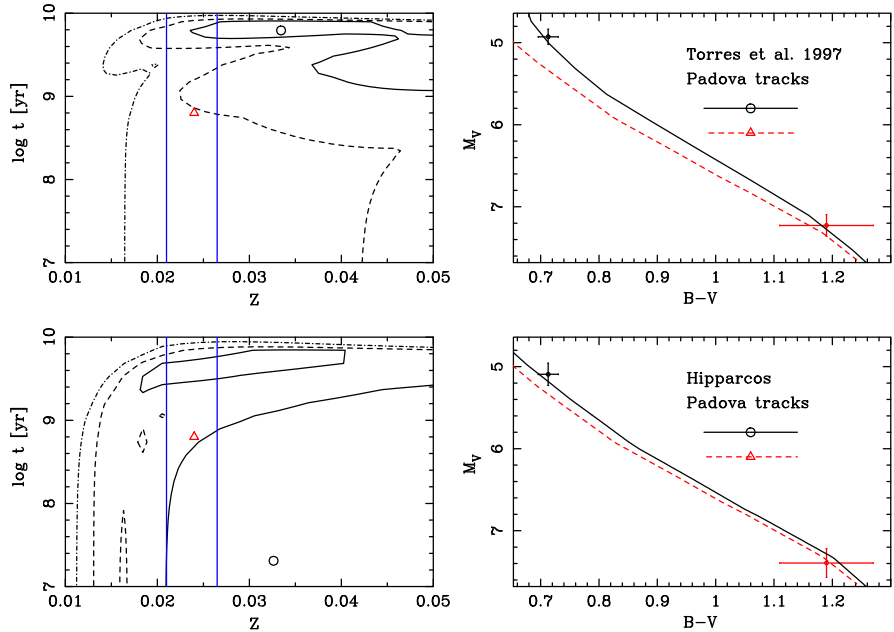}{8. cm}{-90}{45}{45}{-180}{240}
 \caption{\small V818 Tau system: influence of the Hipparcos parallaxes on the contour 
  levels derived from the Padova tracks. The photometric locations of each star in the 
  CMD are from Torres et al. [TSL97a], except the absolute magnitude $M_V$ in the 
  {\it bottom panel} which is derived from the Hipparcos parallax. 
  The best fit isochrones ({\it solid lines}) are defined by Z $=$ 0.033, 
  log t $=$ 9.79 (with Torres et al. parallaxes, {\it upper CMD}) and Z $=$ 0.033, 
  log t $=$ 7.30 (Hipparcos, {\it bottom CMD}), but all the isochrones inside the 
  1$\sigma$ confidence levels are also good fits. 
  {\it Vertical lines} in contour levels diagrams show the observational limits of 
  the Hyades metallicity. The solution (Z $=$ 0.024, log t $=$ 8.80) 
  from Perryman et al. (1998) for the Hyades is also shown for comparison ({\it 
  triangle}). This solution is consistent with the observational metallicity of 
  the Hyades, however, the corresponding isochrone ({\it dashed line}) does not 
  fit the system in the CMD.
   } \label{fig-1}
\end{figure}
\normalsize

\subsection{Results}
Fig. 1 shows that the Hipparcos parallaxes have a strong influence on the solutions 
of V818 Tau in the metallicity-age plane : the solutions obtained with the parallaxes of 
Torres et al. are too old and metal rich. In contrast, the Hipparcos
parallaxes give contours in agreement with the metallicity of the Hyades. 
Hence, the Padova tracks provide 1$\sigma$-contours in agreement with the 
Hyades metallicity only if one takes   
into account the Hipparcos parallax. It is also worth noticing 
that the isochrone corresponding to the Perryman et al. (1998) solution does not 
fit the system in the CMD. \\
The table below briefly summarizes, for the three systems, the results of the 
theoretical simultaneous age--metallicity estimates obtained from isochrone age 
fitting (1$\sigma$ level) taking into account the Hipparcos parallax.

%

\begin{table}
\small
\begin{tabular}{lllllll}
\tableline 
\noalign{\smallskip} 
System         & \multicolumn{2}{c}{Geneva} & \multicolumn{2}{c}{Padova} & \multicolumn{2}{c}{CG92} \\
\noalign{\smallskip}
\tableline
\noalign{\smallskip}
               & Z & log t                  & Z & log t                   & Z & log t               \\ 
\noalign{\smallskip} 
\tableline
\noalign{\smallskip}
\noalign{\smallskip}
51 Tau         & 0.020$^{+0.010}_{-0.008}$ & 8.88$^{+0.22}_{-0.23}$ & 
0.017$^{+0.021}_{-0.005}$ & 8.90$^{+0.15}_{-0.55}$ & 
0.018$^{+0.012}_{-0.006}$ & 8.92$^{+0.23}_{-0.17}$ \\ 
\noalign{\smallskip}
V818 Tau
&                           &                        & 
0.033$^{+0.017}_{-0.015}$ & 7.30$^{+2.50}_{-0.30}$ & 
                          &                        \\ 
\noalign{\smallskip}
$\theta^2$ Tau & 0.027$^{+0.013}_{-0.010}$ & 8.80$^{+0.05}_{-0.09}$ & 
0.027$^{+0.023}_{-0.011}$ & 8.80$^{+0.03}_{-0.11}$ &
0.027$^{+0.003}_{-0.005}$ & 8.88$^{+0.02}_{-0.02}$ \\ 
\noalign{\smallskip}
\tableline
\tableline
\noalign{\smallskip} 
\end{tabular}
\end{table}

For 51 Tau and $\theta^2$ Tau, the 3 sets of isochrones give good fits
in the CMD, in agreement with previous estimates  (Perryman et al.) of
age    (log  t  $=$   $8.80^{+0.02}_{-0.04}$,  from isochrone  fitting
technique with the CESAM  stellar evolutionary code (Morel  1997)) and
metallicity ([Fe/H] $=$ 0.14 $\pm$ 0.05).
The Geneva and CG92  models can not be  tested  with the less  massive
component of V818 Tau. 
It is also interesting to mention that the masses predicted by the 3 sets 
of tracks are in good agreement with the measured individual masses of each 
system but the Padova isochrones can not fit the system V818 Tau in a 
mass-radius diagram (see Lastennet et al. 1999b for more details).



\begin{references}
\reference Andersen, J. 1991, \araa, 3, 91
\reference Charbonnel, C., Meynet, G., Maeder, A., Schaller, G., Schaerer, D.
 1993, \aaps, 101, 415
\reference Claret, A., Gim\'enez, A. 1992, \aaps, 96, 255, [CG92]
\reference ESA, 1997, {\it The Hipparcos and Tycho Catalogues} (ESA-SP 1200)
\reference Fagotto, F., Bressan, A., Bertelli, G., Chiosi, C. 1994, \aaps, 105, 39
\reference Lastennet, E., Lejeune, Th., Westera, P, Buser, R. 1999a, \aap, 341, 857
\reference Lastennet, E., Valls-Gabaud, D., Lejeune, Th., Oblak, E. 
1999b, \aap, accepted (astro-ph/9905273)
\reference Lebreton, Y., G\'omez, A.E., Mermilliod, J.-C., Perryman, M.A.C., 1997, 
{\it Proceedings of the ESA Symposium "Hipparcos - Venice '97"}, ESA-SP-402, p.231
\reference Lejeune, Th., Cuisinier, F., Buser, R. 1997, \aaps, 125, 229
\reference Lejeune, Th., Cuisinier, F., Buser, R. 1998, \aaps, 130, 65
\reference McClure, R.D., 1982, \apj, 254, 606
\reference Morel, P. 1997, \aaps, 124, 597
\reference Perryman, M.A.C., Brown, A.G.A., Lebreton, Y., G\'omez, A., Turon, C., Cayrel de Strobel, G., Mermilliod, J.-C. 
1998, \aap, 331, 81
\reference Torres, G., Stefanik, R.P., Latham, D.W. 1997, \apj, 474, 256, [TSL97a]
\reference Torres, G., Stefanik, R.P., Latham, D.W. 1997, \apj, 479, 268, [TSL97b]
\reference Torres, G., Stefanik, R.P., Latham, D.W. 1997, \apj, 485, 167, [TSL97c]

\end{references}
\end{document}